\documentclass[fleqn,10pt]{SelfArx} 

\usepackage[english]{babel} 

\usepackage{multirow} 

\definecolor{color1}{RGB}{0,0,90} 
\definecolor{color2}{RGB}{0,20,20} 

\usepackage{hyperref} 

\hypersetup{
	hidelinks,
	colorlinks,
	breaklinks=true,
	urlcolor=color2,
	citecolor=color1,
	linkcolor=color1,
	bookmarksopen=false,
	pdftitle={Title},
	pdfauthor={Author},
}

\JournalInfo{Accepted paper presented at the AES 159th convention, Oct 23-25, Long Beach, CA, USA} 
\Archive{This paper was accepted as an Express paper; the complete manuscript was not peer reviewed } 

\PaperTitle{Effects of automotive microphone frequency response characteristics and noise conditions on speech and ASR quality - an experimental evaluation  } 
\ShortTitle{Effects of automotive microphone characteristics on speech and ASR}

\Authors{Michele Buccoli\textsuperscript{1}*, Yu Du\textsuperscript{2}, Jacob Soendergaard\textsuperscript{3} and Simone Shawn Cazzaniga\textsuperscript{1}} 
\FirstAuthor{Buccoli et al.}
\affiliation{\textsuperscript{1}\textit{BdSound S.r.l., Milan, Italy}} 
\affiliation{\textsuperscript{2}\textit{Harman International, Novi -MI, USA}} 
\affiliation{\textsuperscript{3}\textit{HEAD acoustics, Brighton, MI, USA}} 
\affiliation{*\textbf{Corresponding author}: michele.buccoli@bdsound.com} 

\newcommand{\keywordname}{Keywords} 

\Abstract{Upon choosing microphones for automotive hands-free communication or Automatic Speech Recognition (ASR) applications, OEMs typically specify wideband, super wideband or even fullband requirements following established standard recommendations (e.g., ITU-P.1110, ITU-P.1120). In practice, it is often challenging to achieve the preferred bandwidth for an automotive microphone when considering limitations and constraints on microphone placement inside the cabin, and the automotive grade environmental robustness requirements. On the other hand, there seems to be no consensus or sufficient data on the effect of each microphone characteristic on the actual performance. As an attempt to answer this question, we used noise signals recorded in real vehicles and under various driving conditions to experimentally study the relationship between the microphones’ characteristics and the final audio quality of speech communication and performance of ASR engines. We focus on how variations in microphone bandwidth and amplitude frequency response shapes affect the perceptual speech quality. The speech quality results are compared by using ETSI TS 103 281 metrics (S-MOS, N-MOS, G-MOS) and ancillary metrics such as SNR. The ASR results are evaluated with standard metrics such as Word Error Rate (WER). Findings from this study provide knowledge in the understanding of what microphone frequency response characteristics are more relevant for audio quality and choice of proper microphone specifications, particularly for automotive applications.}

\begin{document}

\maketitle 

\tableofcontents 

\thispagestyle{empty} 


\section{Introduction} 
When choosing microphones for automotive hands-free communication or Automatic Speech Recognition (ASR) applications, OEMs typically request wideband, super wideband or even fullband requirements simply following standard recommendations (e.g., ITU-P.1110, ITU-P.1120). In practice, it is often challenging to achieve the specified bandwidth for an automotive microphone when considering limitations and constraints on microphone placement inside the cabin, and the automotive grade environmental robustness requirements. Moreover, there seems to be no consensus or sufficient data on the effect of each microphone characteristic and mounting location on the actual performance perceived by end users. 

Some related works linking automotive microphone specifications to user experiences can be found in the literature; but results are limited. A first investigation was conducted in \cite{Du:2019dg}, where the authors discuss the correlation between three automotive microphones and perceived user experience, measured as Speech Intelligibility Index (SII) and subjective listening.

The work was extended in \cite{Du:2023dg} including both objective and subjective speech intelligibility and quality. In \cite{Maver:2024dg} the authors presented an acoustic front end (AFE) composed of source separation and noise reduction and evaluated their performance with objective metrics (Signal-to-Noise Ratio, SNR, and Signal-to-Interferent Ratio, SIR) across four different automotive microphone types and placements in an automotive environment. 

These works were based on selected types of automotive microphones, hence the research space was limited to characteristics associated with those microphones. Previous studies did not show the general trends of how variations in microphone characteristics affect the speech and ASR quality.

Attempting to answer this question, this paper focuses on how variations in microphone bandwidth and frequency response (FR) shapes affect not only the perceptual speech quality on a Human-to-Human (H2H) call, but also the ASR performance for Human-to-Machine (H2M) interactions, under different driving conditions. 

The speech quality results of the audio files are compared using mainly ETSI TS 103 281 metrics (S-MOS, N-MOS, G-MOS) and ancillary metrics such as ETSI TS 103 558 metrics (listening effort) and SNR [4, 5]. By doing so, we evaluate how variations in microphone FR characteristics affect the speech quality and noise transmission quality as captured by the microphones at different noise conditions. 

Secondly, we also apply the same set of audio files to an open-source speech recognition engines to compare the resulting Word Error Rate (WER).  This allows us to check for any correlation between the ASR performance and changes in microphone FR specifications under different driving conditions.

Findings from this study provides knowledge in the understanding of what microphone frequency response characteristics are more relevant for the audio quality and the choice of proper microphone specifications, particularly for automotive applications.


\begin{figure}[!t]\centering 
	\includegraphics[width=\columnwidth]{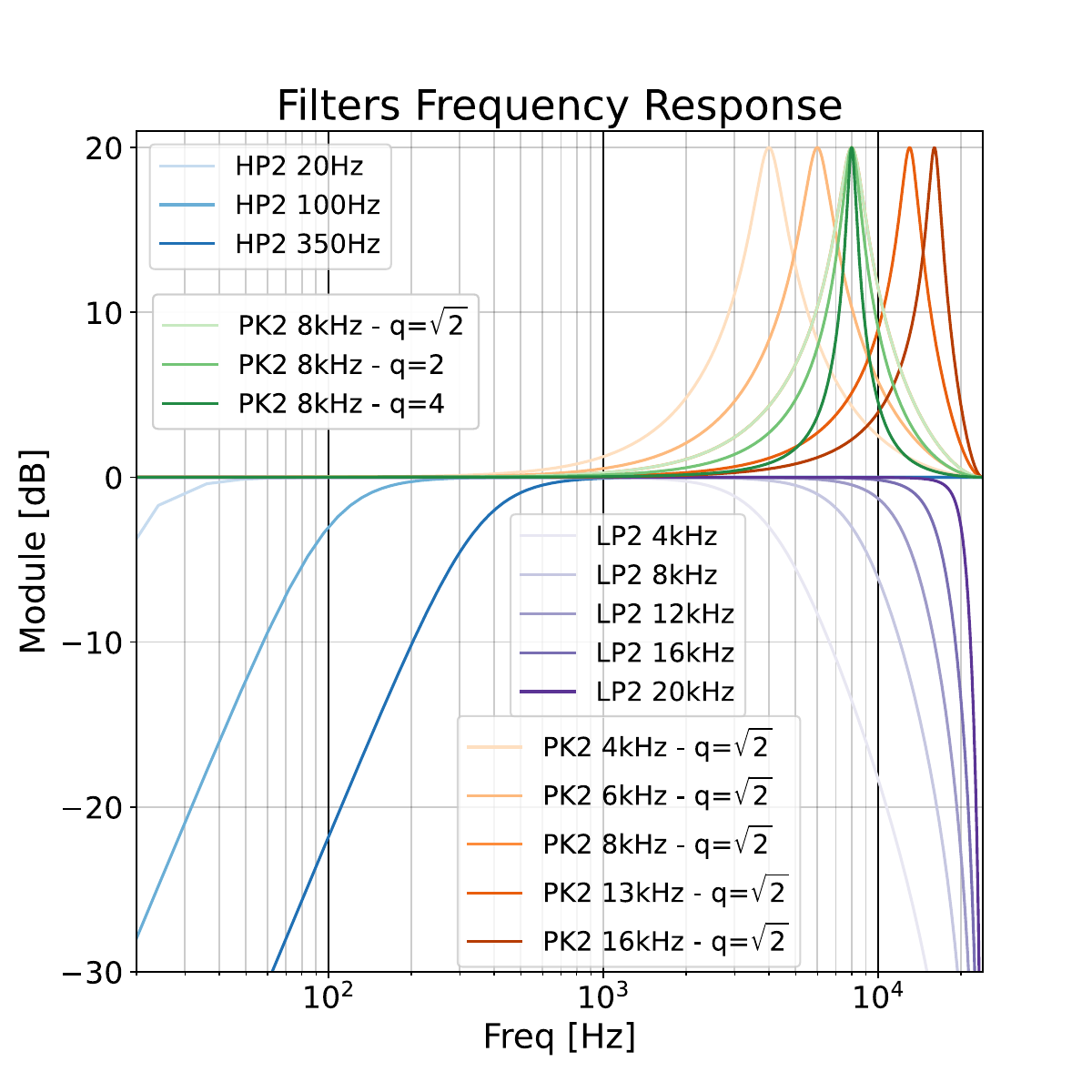}
	\caption{Representation of microphone’s FR characteristics emulated in this study.}
	\label{fig:FRchar}
\end{figure}

\section{Recording simulations}
The analysis has been carried out over simulated recordings $x(n)$, that are calculated as in Equation \ref{eq:formulation} by convolving a clean ITU-T P.501  \cite{ITU:P501:2025} compliant speech $s(n)$ with a Car Impulse Response $h(n)$, added to a real Car Background Noise $v(n)$ and finally passed through a cascade of digital filters $f(\cdot)$ simulating various types of spectral characteristics of possible designs of automotive microphones: 
\begin{equation}
x(n) =f( s(n) \star h(n) + v(n))
\label{eq:formulation}
\end{equation}

\subsection{Stimulus and recordings}
The speech stimulus is described in ETSI TS 103 281  Annex E \cite{ETSI:TS103281:2019}. It is constructed using 20 American English Harvard sentences, which are phonetically balanced \cite{IEEE:297:1969} and consist of 1 second of leading silence and 1 second of trailing silence, for a total sentence length of 4 seconds. It uses multiple different male and female talkers. Total stimulus length is 80 seconds.

Car Impulse Responses are extracted from recordings of Head and Torso Simulators (HATS) with artificial mouth placed in the driver position and captured by a high-quality measurement microphone (i.e., nearly flat frequency response in the full audio band) placed at the location of the in-vehicle hands-free microphone position – typically the driver’s side visor or overhead consol position . Three car types are used in this study. They are a mid-size Sedan, a compact and a subcompact SUV.  

Background noises are recorded using the same high-quality measurement microphone and during road tests, following the guidelines outlined in the Annex D of the ITU P.1100 standard \cite{ITU:P1100:2019}, namely: idle (stationary vehicle with low fan), city (driving at 60 kmph with medium fan) and highway (120 kmph at low fan). In this paper, we do not include the fourth and fifth scenario, which has driving at 120 and 180 kmph, respectively.

\subsection{Microphone FR simulations}
Microphone FR characteristics are typically determined by the signal bandwidth and resonance behavior, which are simulated with a cascade of digital filters in this study. The single stages, all designed as second-order bilinear transform filters, are represented in Fig. \ref{fig:FRchar} and listed in Table \ref{tab:filters}.   

The bandwidth is defined by a low-pass (LP2) and a high-pass (HP2) filter. Selections of corner frequencies for LP2 and HP2 cover common bandwidth requirements including narrowband (4 kHz), wideband (8 kHz), super-wideband (up to 16 kHz) and fullband (20 kHz). 

A second order peak filter (PK2) with a fixed amplitude of 20 dB is used to simulate the resonance behavior typically exhibited by a commercial microphone module due to its acoustic designs of the porthole, housing and face grille. The peak filter is positioned between 4 kHz and 16 kHz with a quality factor (q-factor) of 1.414, 2 and 4.  

Note that, if all possible parameter combinations are considered in Table \ref{tab:filters}, a total of 225 microphone FR shapes can be generated. Based on practical use cases and limitations, we strategically selected 113 microphone FR shapes in this study.

\begin{table}[t]
	\caption{Filter stages for microphone characteristics settings.}
	\centering
	\begin{tabular}{|l|l|l|}
		\toprule
		Filter & Frequency [Hz] & q-factor \\
		\midrule
        HP2 & 20, 100, 350 &	0.707 \\
        LP2 & 4k, 8k, 12k, 16k, 20k	& 0.707 \\
        PK2 & 4k, 6k, 8k, 13k, 16k	& 1.414, 2, 4 \\            
		\bottomrule
	\end{tabular}
	\label{tab:filters}
\end{table}

\section{Results and discussions}
A total of 1017 (113 microphone FR by 3 car types by 3 noise types) speech files are generated for subsequent sound quality evaluation following ETSI TS 103 281 metrics (S-MOS, N-MOS, G-MOS). S-MOS specifically evaluates the speech components of the signal, while N-MOS evaluates the intrusiveness of the noise components of the signal. G-MOS is the global (or overall) quality impression of the signal. All the MOS scores are on a scale of 1 ("very distorted” for S-MOS and “very intrusive” for N-MOS) to 5 ("not distorted” for S-MOS and “not noticeable” for N-MOS), with qualitative descriptions found in ITU-T P.835 \cite{ITU:P835:2003}. The MOS scores are computed using a commercially available software package. Note that since G-MOS values are calculated using a random forest regression from S-MOS and N-MOS values, there is strong correlation between them, and in the interest of brevity, G-MOS results are not reported in this paper. A listening effort metric following ETSI TS 103 558 \cite{ETSI:TS103558:2021} is also computed and used to gauge speech intelligibility and to see if it differs from speech quality for certain microphone bandwidth configurations. Additionally, the A-weighted SNR is calculated for each scenario and used to verify conclusions.

For each sentence of the speech file, a data point is generated, resulting in 20 data points for the same study case (representing 1 microphone, 1 car type and 1 noise type) so that necessary statistical analysis can be applied.   

The same files are also analyzed with the popular ASR engine Whisper on its tiny model \cite{Radford:2023dg}. The ASR performance is estimated with the Word Error Rate (WER) metric. The analysis has produced one datapoint for each combination of microphone parameters, car and noise scenarios, aggregating the results for the 20 sentences processed with each of these combinations.

The analysis on the effects of automotive microphones characteristics on speech quality and recognition is conducted by visually inspecting boxplots generated by the distribution of the data points. ANOVA test is also conducted to evaluate the statistical significance of the underlying results.

\begin{figure}[!t]\centering 
	\includegraphics[width=\columnwidth]{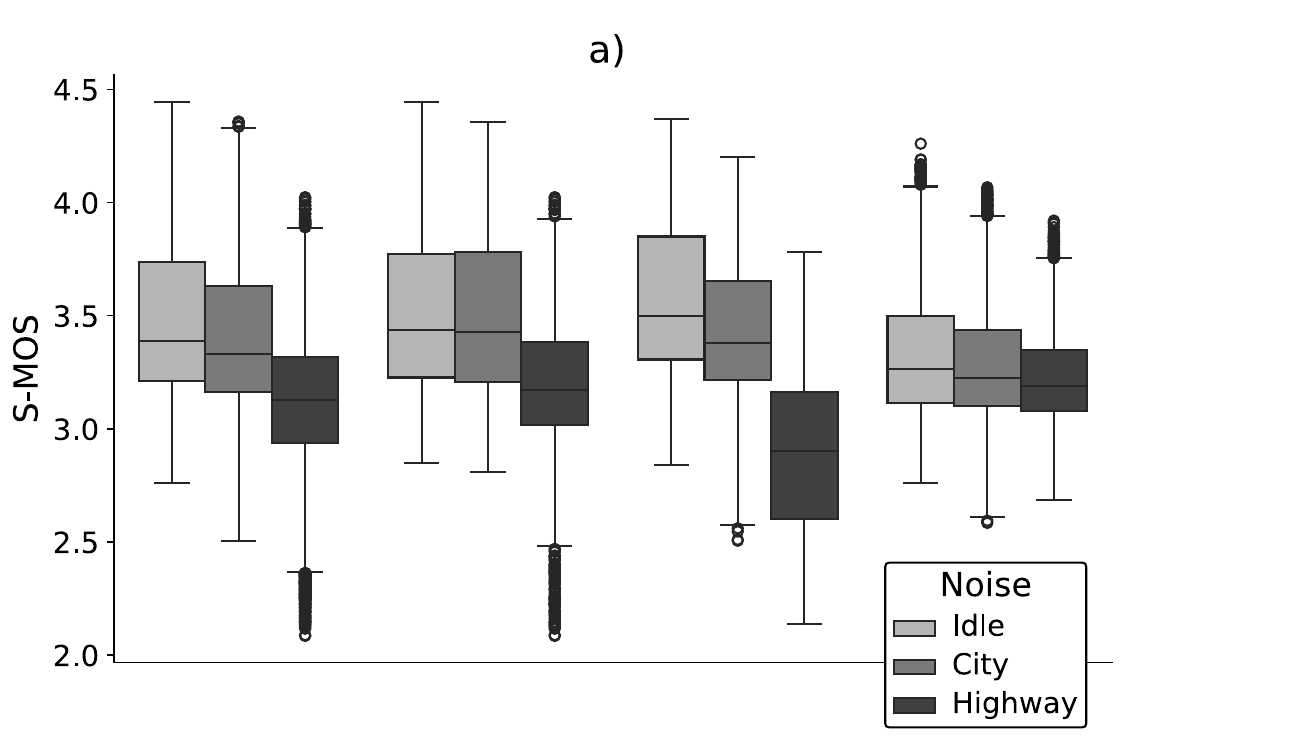}
	\includegraphics[width=\columnwidth]{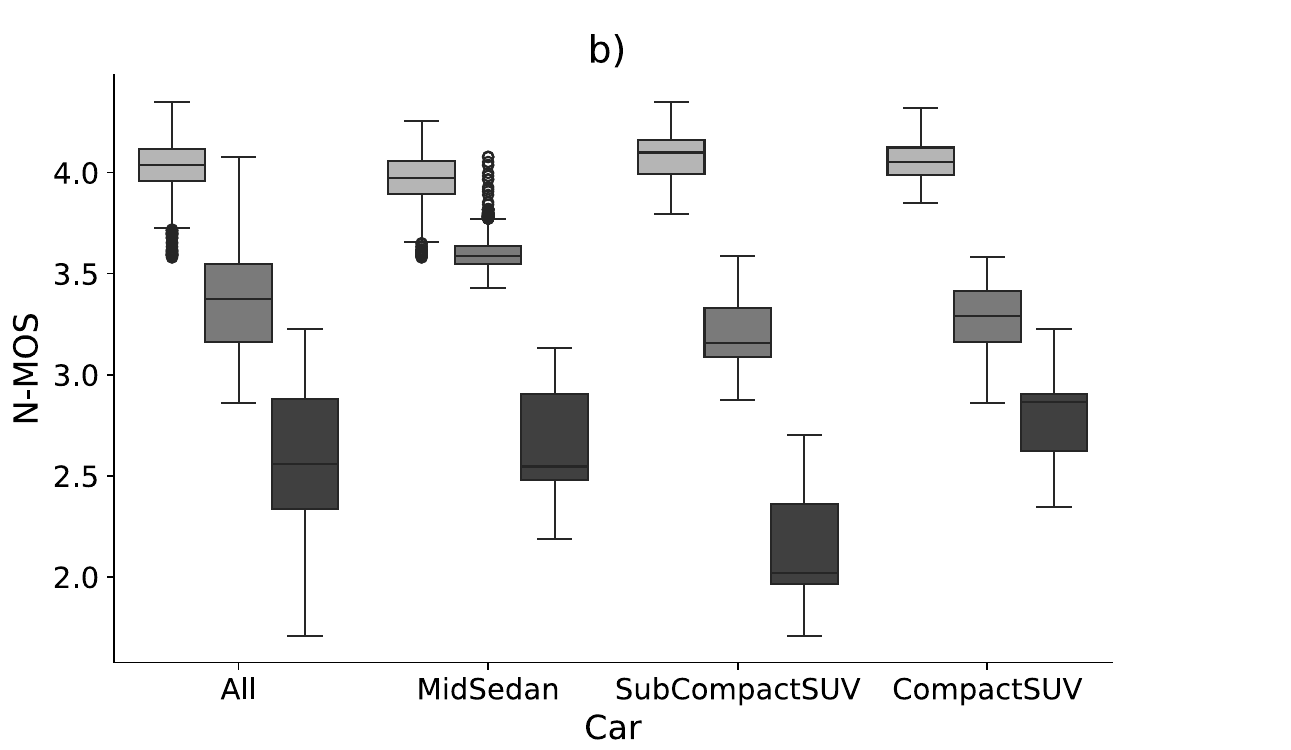}
	
    \caption{S-MOS (a) and N-MOS (b) values separated by noise and car types.}
	\label{fig:SNMOSall_huenoise}
\end{figure}

\subsection{Effects of car and noise types}
To investigate how the simulated sound quality correlates to the noise and car types, S-, N-, and G-MOS values are calculated using data separated by three noise and three car types but with all microphone FR simulation cases included.

\begin{figure}[!t]\centering 
	\includegraphics[width=\columnwidth]{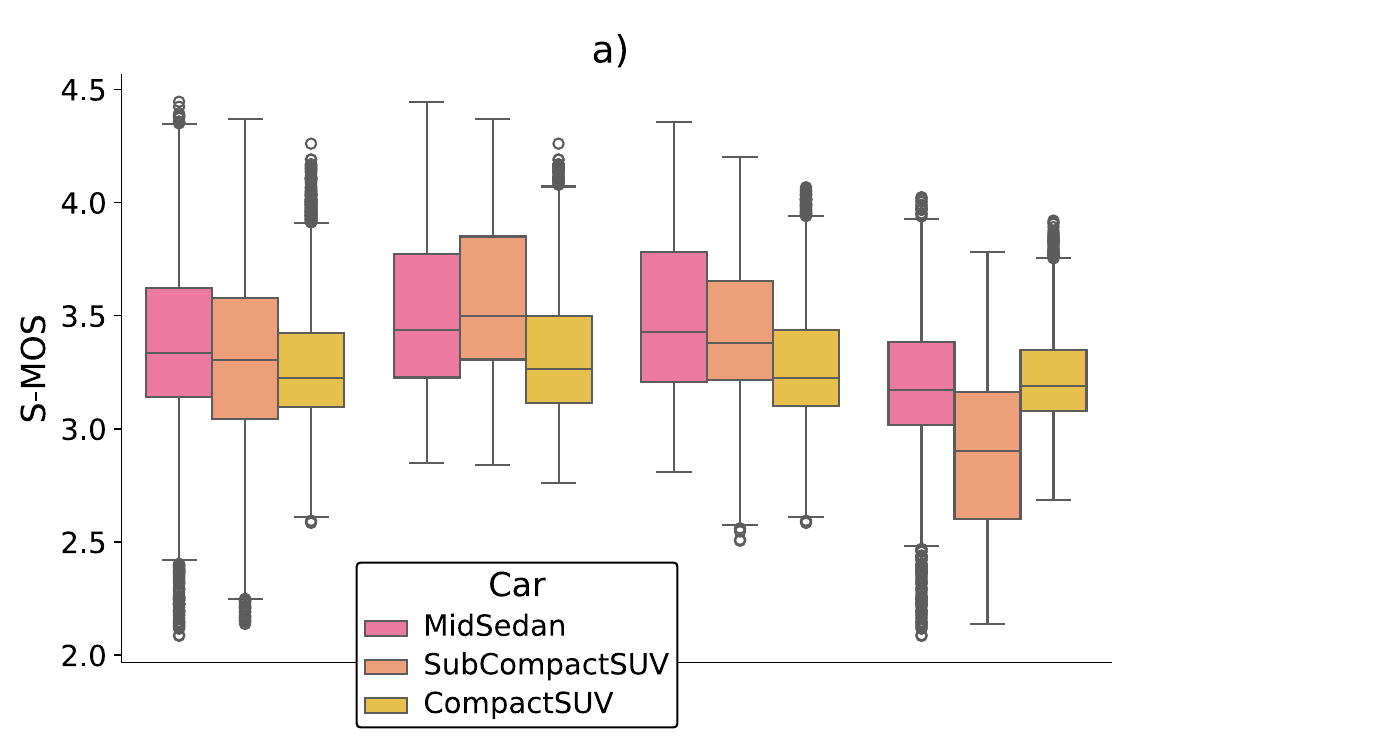}
	\includegraphics[width=\columnwidth]{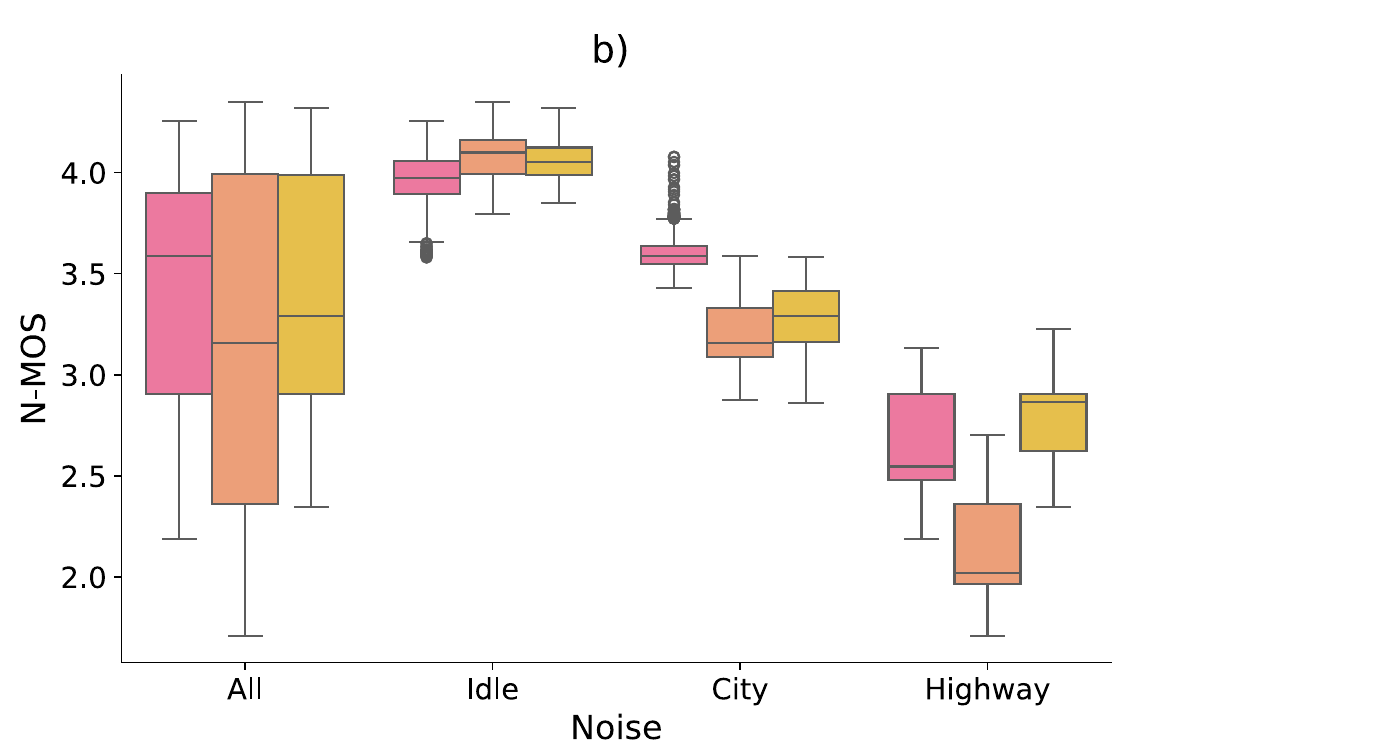}
    \caption{S-MOS (a) and N-MOS (b) values separated by car and noise type.}
	\label{fig:SNMOSall_huecars}
\end{figure}

\begin{figure}[b]\centering 
	\includegraphics[width=\columnwidth]{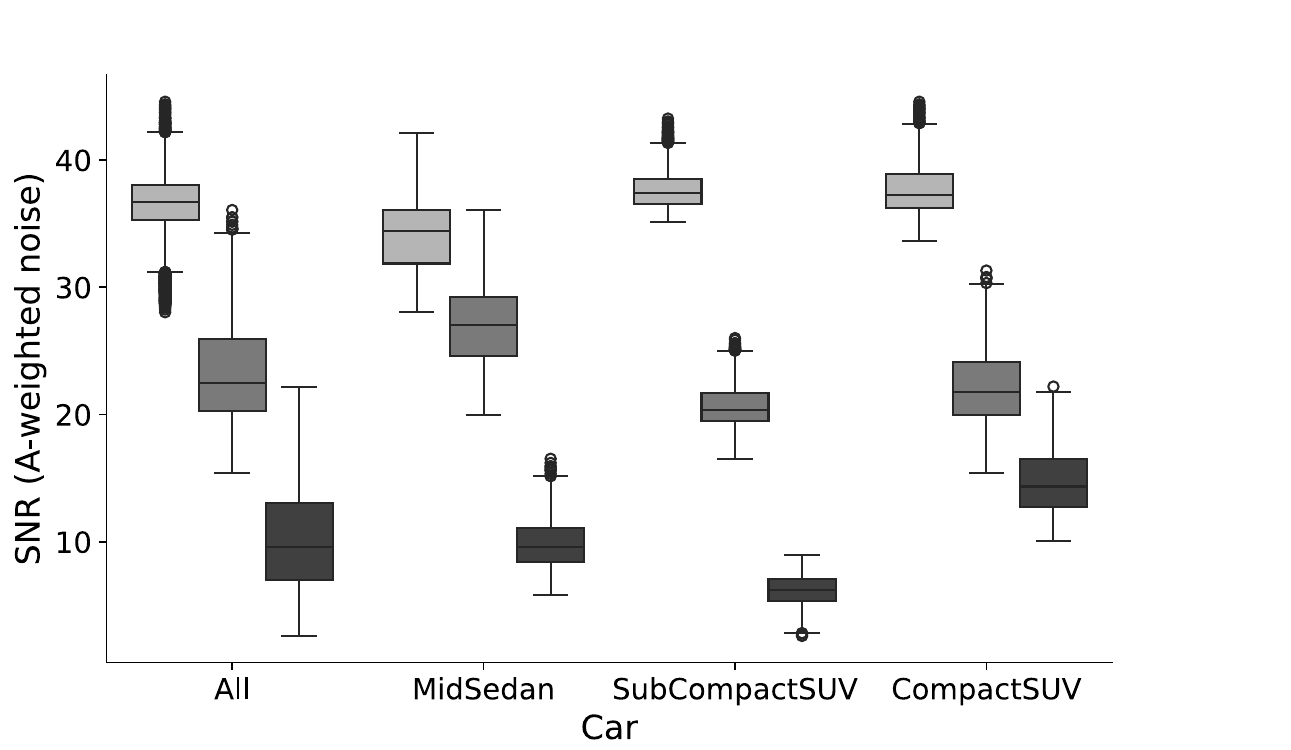}
	\caption{A-weighted SNR as a function of noise (hue) and car type. Legend is the same of Fig. \ref{fig:SNMOSall_huenoise}}
	\label{fig:SNRall}
\end{figure}

Figures \ref{fig:SNMOSall_huenoise}a and \ref{fig:SNMOSall_huenoise}b show box plots of the S-MOS and N-MOS values, respectively, calculated from all data points separated only by the noise type (i.e., idle, city and highway) and for each car type. A trend showing decreasing speech quality with increasing background noise level can be seen clearly, and confirmed by statistical significance test with p-values close to 0 for ANOVA on both S-MOS and N-MOS. This noise degradation trend is observed when the MOS values are compared for each car type (i.e., mid-size Sedan, compact and subcompact SUV). Also in this case, the corresponding p-values are close to zero, validating that S-MOS and N-MOS values vary with noise type.

Similarly, box plots of the S-MOS and N-MOS values separated by the car type are shown in Fig. \ref{fig:SNMOSall_huecars}. Unlike the noise type, the car type seems to affect MOS values without a clear trend. When all data points separated only by car types are examined, S-MOS for all three car types are very close to each other (Fig. \ref{fig:SNMOSall_huecars}a). The median value of the N-MOS (Fig \ref{fig:SNMOSall_huecars}b) does show some discrepancies across car types. This is confirmed by the statistical analysis, where we observe a p-value close to 0, with a low corresponding F-statistics, indicating a smaller difference between groups. In particular, the subcompact SUV consistently gives lowest S- and N-MOS values among all three car types. However, the A-weighted SNR results in Fig. \ref{fig:SNRall} also indicate that the subcompact SUV has the lowest SNR value under the highway noise condition, which implies that the low performance of the subcompact SUV under the highway noise condition is again associated more with the noise level than the car type. 

Based on the above observations and for the noise and car types included in this study, we conclude that the noise type (i.e., level) has a significant effect on the speech quality. On the other hand, MOS values do not seem to be affected by the car type with a clear trend. For simplicity and straightforward data processing, further plots and analyses in this paper will combine data for all car types together.

\begin{figure}[!t]\centering 
	\includegraphics[width=\columnwidth]{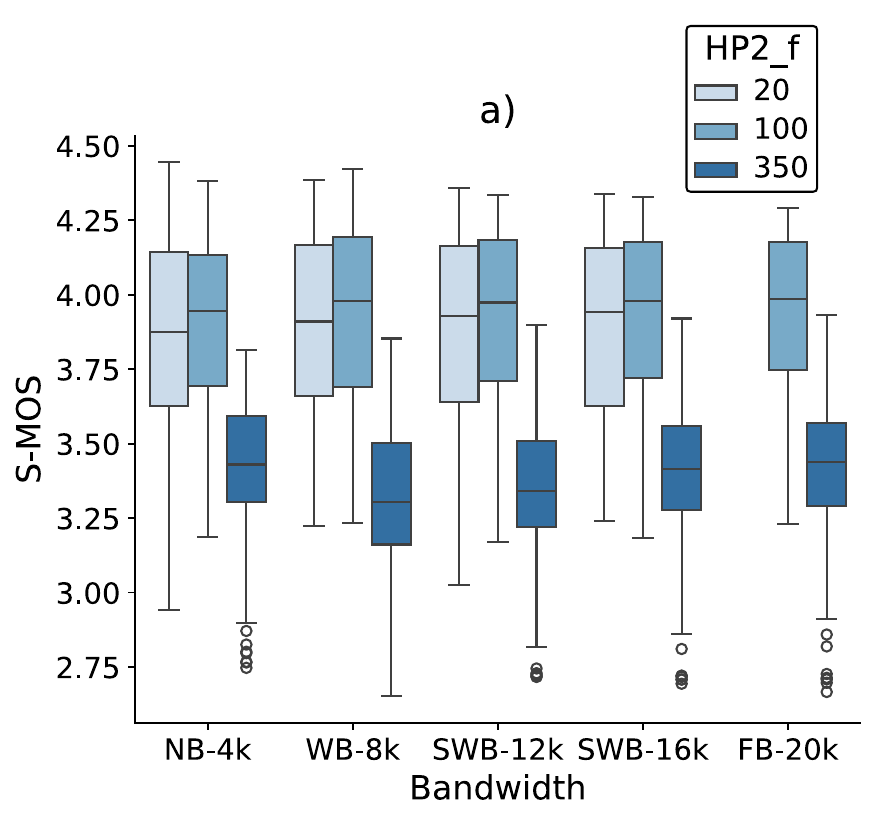}
	\includegraphics[width=\columnwidth]{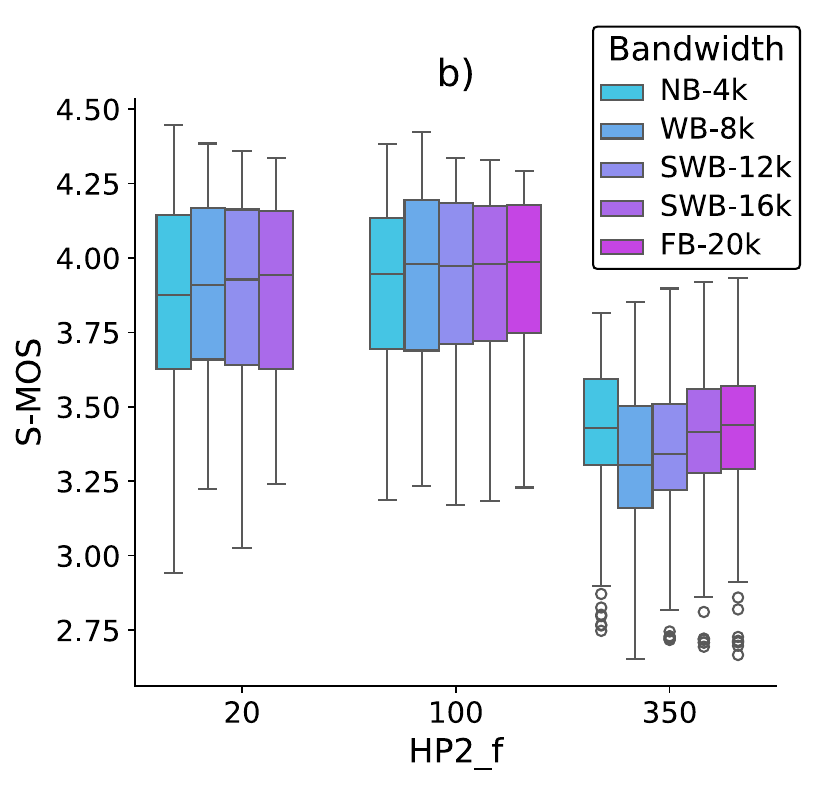}
    \caption{S-MOS values versus (a) high and (b) low cut-off frequencies.}
	\label{fig:SMOS_bw_cutoff}
\end{figure}

\subsection{Effects of microphone signal bandwidth}
The microphone bandwidth is defined as the frequency range between the cut-off (corner) frequencies of the second order high-pass and low-pass filters (HP2 and LP2). To evaluate the bandwidth effects alone, the microphone is assumed to have a flat FR shape in the entire bandwidth without including the peak filters (PK2). Combinations of three HP2 and five LP2 corner frequencies, as detailed in Table 1. Cut-off frequencies at the higher end of the bandwidth represent the narrowband, wide wideband, two super wideband and full-band cases, respectively.

The S-MOS values versus lower and higher end bandwidth cut-off frequencies are plotted in Fig. \ref{fig:SMOS_bw_cutoff}, which are calculated using all data points regardless noise and car types. Two observations may be made from Fig. \ref{fig:SMOS_bw_cutoff}. Firstly, with the same high cut-off frequency (indicated by NB/WB/SWB/FB), low cut-off frequency at 20 and 100 Hz result in similar S-MOS values and are apparently higher than the 350 Hz case. Secondly, with the same low cut-off frequency, the higher limit of the bandwidth does not seem to matter much as the S-MOS values are very similar to each other. A closer look at Fig. \ref{fig:SMOS_bw_cutoff}(b) may imply that, with a high lower end cut-off frequency at 350 Hz, the 4kHz narrowband case gives slightly higher S-MOS value than other bandwidth options. The p-value score is close to 0, enforcing this observation, but the corresponding F-statistic is low, showing little difference among groups.

\begin{figure}[!b]\centering 
	\includegraphics[width=\columnwidth]{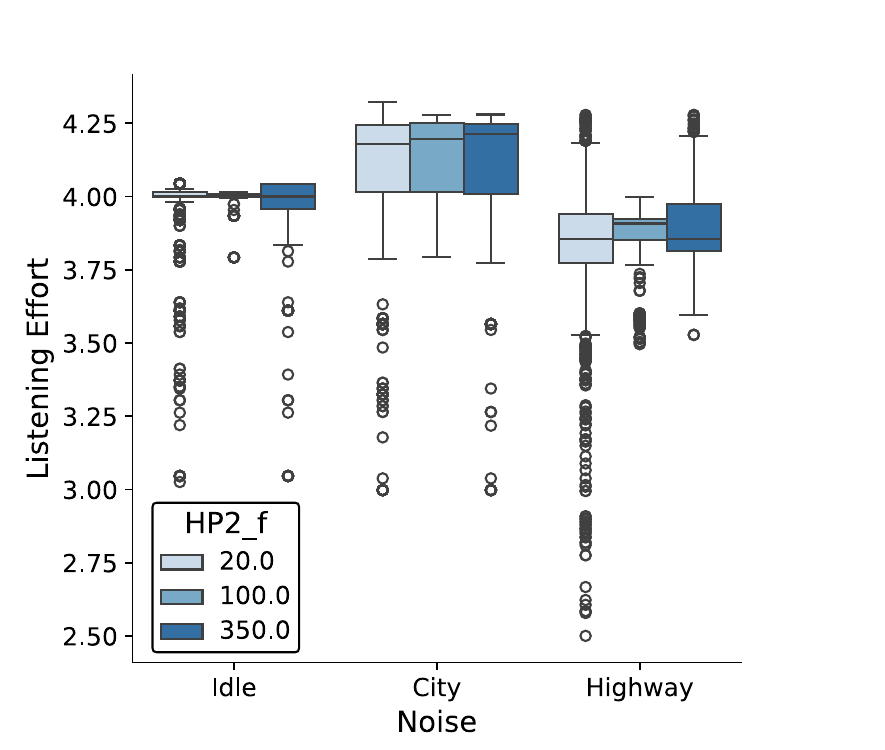}
	\caption{Listening effort evaluation for different bandwidths at varying noise type.}
	\label{fig:LE_bw}
\end{figure}

Note that the 350 Hz cutoff frequency at the low end of the bandwidth is included in this study for comparison, typical low end bandwidth cutoff frequencies are 20, 50 or 100 Hz (see for example ITU-P.1100, 1110, and 1120). 

\begin{table}[!t]
	\caption{ANOVA results for S-MOS and N-MOS at different corner frequencies.}
	\centering
	\begin{tabular}{|l|l|l|l|}
		\toprule
        Subset	& Metric &	p &	F \\
		\midrule
        \multirow{2}{*}{LP2\_f with all HP2\_f} & S-MOS & 0.755 & 0.47 \\        
                                               & N-MOS	&0.516 &	0.81 \\
        \midrule                                               
        \multirow{2}{*}{LP2\_f with HP2\_f = 350 Hz} &	S-MOS &	0.008 &	3.49 \\
                                                    &	N-MOS &	0.923 &	0.23 \\
        \midrule
        \multirow{2}{*}{HP2\_f with all LP2\_f}	& S-MOS &	0	& 1174 \\
                                                &	N-MOS &	0 &	30.76 \\
        \midrule
        \multirow{2}{*}{HP2\_f with LP2\_f = 8kHz}	& S-MOS &	0 &	242.66 \\
	                                                & N-MOS &  0.002 & 6.04 \\
		\bottomrule
	\end{tabular}
	\label{tab:ANOVA}
\end{table}

Table \ref{tab:ANOVA} shows the p-values for S-MOS and N-MOS at different LP2 frequencies for all HP2 frequencies, and for the specific 350 Hz HP2 frequency and at different HP2 frequencies for all LP2 and the specific 8kHz frequency. The specific frequencies were chosen as those with the lowest p-values and highest F-statistics. It is observable that variations on N-MOS are either not statistically significant or rather moderate with respect to the S-MOS counterpart. For this reason, the N-MOS plots are not reproduced here for brevity.

Excluding the 350 Hz case and combining the information obtained from Figs. \ref{fig:SMOS_bw_cutoff} and Table \ref{tab:ANOVA}, it may be concluded that the microphone signal bandwidth does not have significant effects on the speech audio quality measured by S-MOS. Another support for this conclusion is demonstrated in Fig. \ref{fig:LE_bw} that shows the MOS values focusing on the listening effort based on ETSI TS 103 558 \cite{ETSI:TS103558:2021}. While the ANOVA test results with low p-values, hence suggesting different groups, a visual inspection of Fig. \ref{fig:LE_bw} indicates that the listening effort values are about the same under each noise type regardless of the lower cutoff frequency values.

\begin{figure}[!t]\centering 
	\includegraphics[width=\columnwidth]{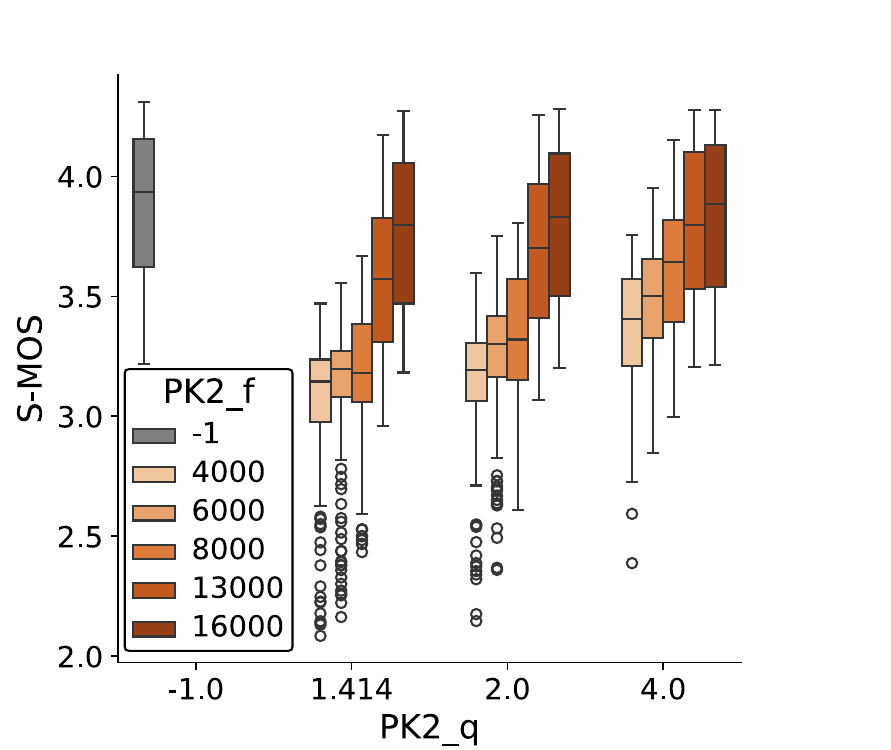}
	\caption{S-MOS values as a function of resonance peak quality, calculated for full-band case (-1 in the legend means no response peak).}
	\label{fig:SMOS_PK2_q}
\end{figure}

\subsection{Effects of microphone FR peak}
Due to practical design and installation limitations, an automotive microphone usually presents a resonance peak in the audio band. It is of interest to understand how the resonance peak location and quality factor affect the audio quality. From ANOVA tests, we verify that variations of these two factors do not affect the N-MOS in a statistically significant way, and therefore N-MOS is not shown in this section.

The S-MOS values showing the peak location and quality factor effects with a full-band microphone are plotted in Figures 7 and 8. We observe in Fig. \ref{fig:SMOS_PK2_q} that lower peak frequency leads to lower S-MOS values. It also seems that S-MOS values are similar for resonance peaks at or below 8 kHz. Furthermore, it may be concluded that if a resonance peak in the microphone FR is inevitable, it should be better pushed above 10 kHz. 

\begin{figure}[!t]\centering 
	\includegraphics[width=\columnwidth]{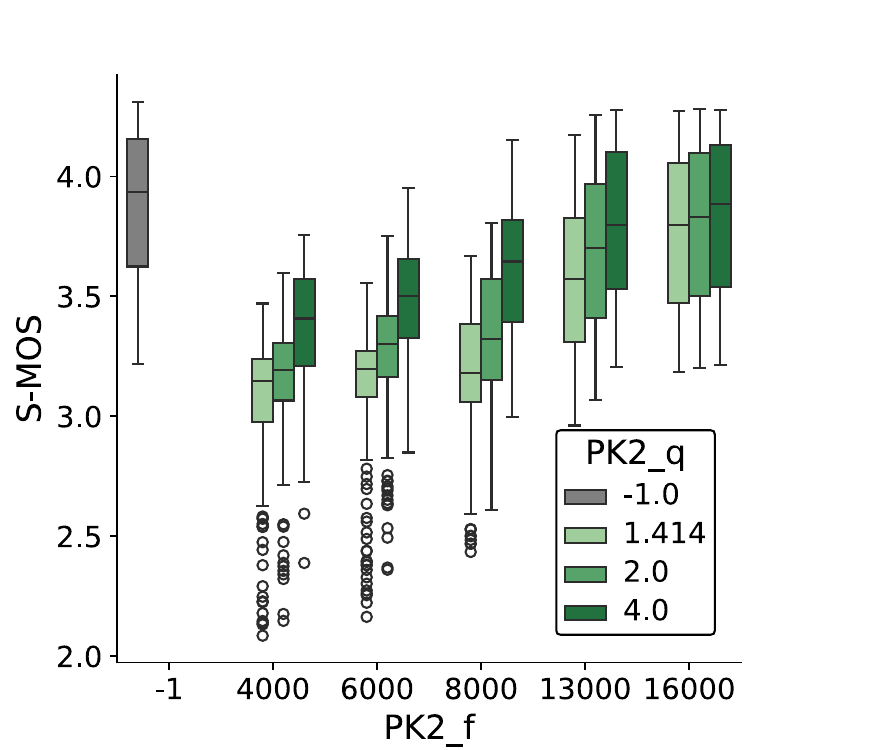}
	\caption{S-MOS values as a function of resonance peak frequency, calculated for full-band case (-1 in the legend means no response peak applied).}
	\label{fig:SMOS_PK2_f}
\end{figure}

With fixed peak frequency, from Fig. \ref{fig:SMOS_PK2_f} we observe that higher quality factor results in better S-MOS performance. Note that in this study higher peak quality factor corresponds to narrower peak bandwidth since the peak amplitude is fixed at 20 dB. Therefore, the quality factor effect should be reevaluated by varying the amplitude while fixing the peak bandwidth, which should be a future task.

\begin{figure}[!t]\centering 
	\includegraphics[width=\columnwidth]{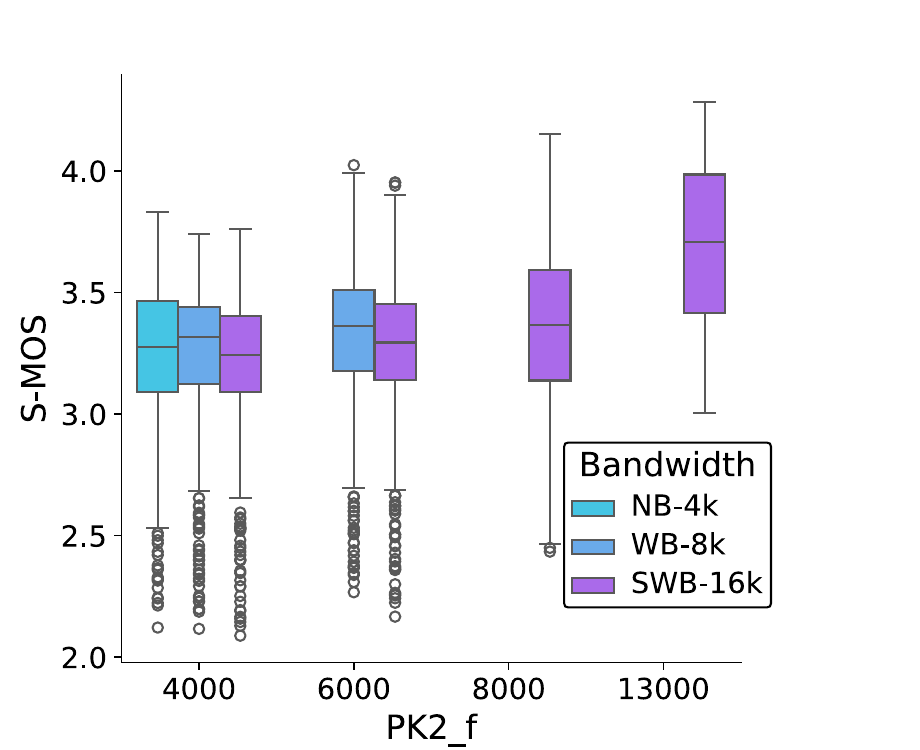}
	\caption{S-MOS values as a function of resonance peak frequency, calculated for narrowband, wideband and 16kHz super wideband cases. }
	\label{fig:SMOS_pk2f_bw}
\end{figure}

In Fig. \ref{fig:SMOS_pk2f_bw} we show that effects of the resonance peak on the audio speech quality are similar for different bandwidth settings.  

\begin{figure}[t]\centering 
	\includegraphics[width=\columnwidth]{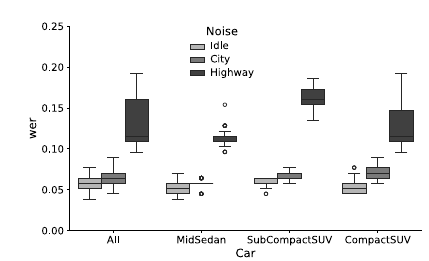}
	\caption{WER for different types of cars, including all aggregated. }
	\label{fig:wer_noise}
\end{figure}

\begin{figure}[t]\centering 
	\includegraphics[width=\columnwidth]{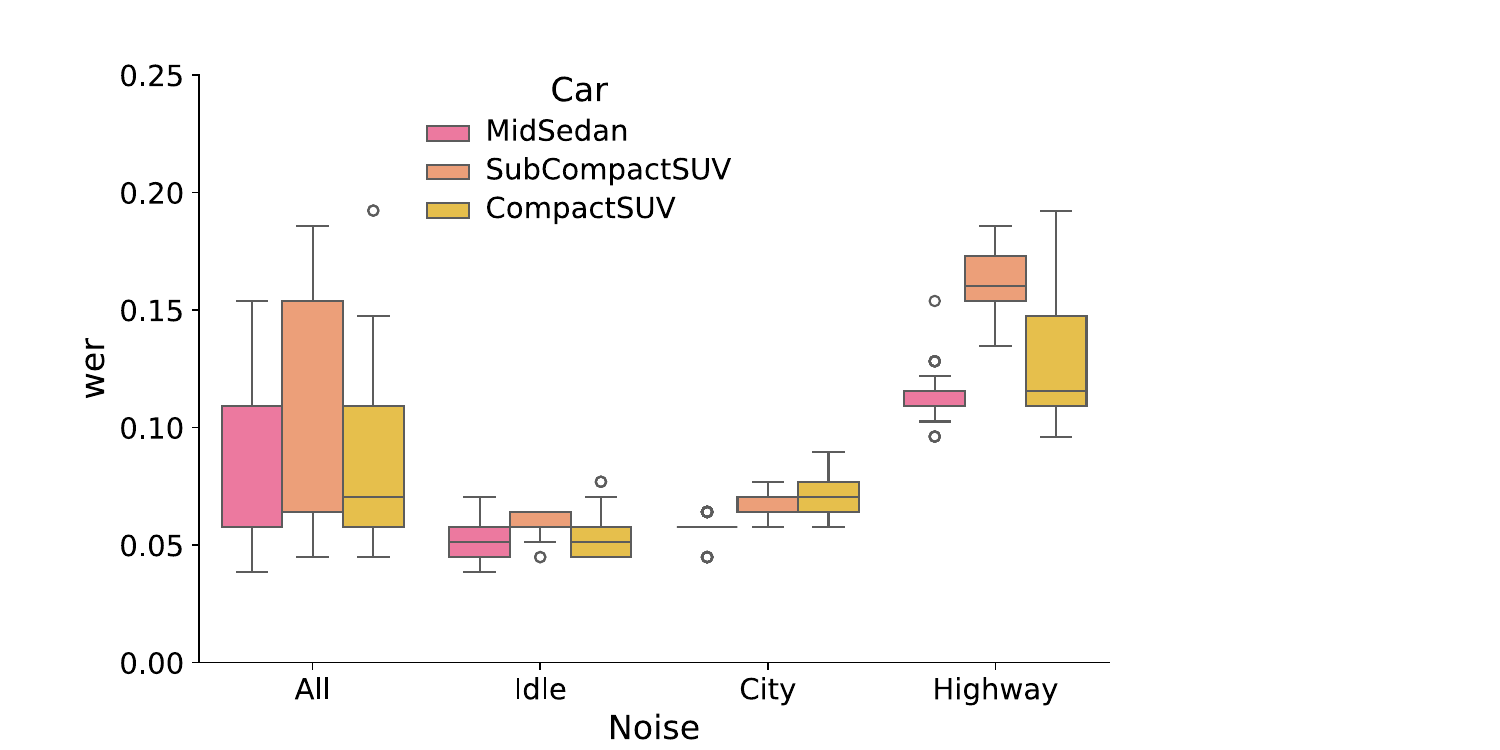}
	\caption{WER for different types of noise, including all aggregated.}
	\label{fig:wer_car}
\end{figure}

\subsection{Effects of microphone FR on ASR}
We have repeated the above analysis for ASR evaluation. From a preliminary statistical analysis, microphone FR characteristics do not affect the ASR performance. There are two possible reasons for this. A first reason is that the ASR engine is robust to several alteration of a speech signal, including changes of FR characteristics. A second reason is that the speech stimulus was present in the training set (that is not public) of the ASR engine, therefore causing overfitting.

The main cause of different performance on ASR is the noise type, with higher noise type negatively affecting the performance on WER, as shown in Figure \ref{fig:wer_noise}. While also car type affects the WER, as shown in Figure \ref{fig:wer_car} there is not a clear trend. Both noise and car effects correspond to a p-value equal to 0, with a higher difference for noise variations than for car variations.

\section{Conclusions and future work}
This study presents a preliminary investigation into how the frequency response characteristics of a selection of microphones affect a subset of speech quality metrics and automatic speech recognition performance. Our work establishes a framework for analyzing microphone FR directly from real-world recordings and provides an evaluation platform aimed at generating data to support informed decision-making for microphone selection in automotive environments. The primary outcomes of this analysis are (1) noise type and levels are the most relevant factor to affect speech quality and recognition, (2) microphone bandwidth has little influence on speech quality (or intelligibility), (3) S-MOS degrades when the lower cutoff (HP2) moves higher than 100Hz, (4) naturally occurring resonances in microphone response should be pushed as high as possible and with a narrow peak width (i.e., high q factor) for best possible quality outcome, (5) ASR performance appears unaffected by all microphone factors.

Our research, while foundational, reveals several avenues for future investigation, as numerous factors can influence speech quality and recognition.

The scope of our initial study was limited to three specific car models and their associated noise recordings. Our analysis shows that car type influences both speech quality and recognition, therefore we plan to broaden our analysis to include a wider range of vehicle models. This will allow us to analyze the effect of cars’ objective characteristics, such as size, class or RT60 on speech quality and recognition. We also plan to recreate all possible combinations of car models and driving noises, which will allow us to more effectively decouple the influence of the driving conditions from the car model itself.

Furthermore, our analysis relies on aggregate statistics derived from 20 sentences spoken by various speakers. It would be valuable to investigate how speaker-specific characteristics, such as pitch frequency, interact with microphone characteristics to affect speech quality on an individual level. In this study, we employed second-order filters to simulate different microphone characteristics, but we intend to explore whether filters with other design features would have different effects on speech quality and ASR performance, for instance, low-pass and high-pass filters of lower or higher orders and peak filters with amplitudes lower and higher than 20 dB.

This work also utilized a popular, yet general-purpose ASR engine. We intend to expand our metrics by evaluating performance with different engines, including those designed for automotive applications. Finally, since microphone signals are often processed by an acoustic front-end (AFE) to reduce noise and enhance speech quality, we plan to extend this work by processing the different microphones with commercially-available AFEs. This will allow us to determine whether these systems can mitigate or even exacerbate the effects of a microphone's inherent characteristics.

\section{{Acknowledgments}}
The study presented in this paper was a collaborative effort by the speech subcommittee of the AES automotive audio technical committee. The authors would like to thank all committee members who participated in discussions and shared their expertise and guidance to support this effort.






\phantomsection
\bibliographystyle{unsrt}
\bibliography{biblio.bib}


\end{document}